\documentclass[pdflatex,sn-mathphys-num]{sn-jnl}%

\usepackage{graphicx}%
\usepackage{multirow}%
\usepackage{amsmath,amssymb,amsfonts}%
\usepackage{amsthm}%
\usepackage{mathrsfs}%
\usepackage[title]{appendix}%
\usepackage{xcolor}%
\usepackage{textcomp}%
\usepackage{booktabs}%
\usepackage{algorithm}%
\usepackage{algorithmicx}%
\usepackage{algpseudocode}%
\usepackage{listings}%
\usepackage{subcaption}
\usepackage[capitalise]{cleveref}
\usepackage{tabularx}
\usepackage{comment}
\newcolumntype{L}{>{\raggedright\arraybackslash}X}

\theoremstyle{thmstyleone}%

\theoremstyle{thmstyletwo}%

\theoremstyle{thmstylethree}%
\begin{document}

\title{High-level Stream Processing: A Complementary Analysis of Fault Recovery}

\author*[1]{\fnm{Adriano} \sur{Vogel}}%

\author[1]{\fnm{Sören} \sur{Henning}}%

\author[2]{\fnm{Esteban} \sur{Perez-Wohlfeil}} %

\author[2]{\fnm{Otmar} \sur{Ertl}}%

\author[3]{\fnm{Rick} \sur{Rabiser}}%

\affil[1]{\orgdiv{JKU/Dynatrace Co-Innovation Lab, LIT CPS Lab}, \orgname{Johannes Kepler University Linz}, \orgaddress{\city{Linz}, \country{Austria}}}

\affil[2]{\orgdiv{Dynatrace Research}, \orgaddress{\city{Linz}, \country{Austria}}}

\affil[3]{\orgdiv{LIT CPS Lab}, \orgname{Johannes Kepler University Linz}, \orgaddress{\city{Linz}, \country{Austria}}}

\abstract{
Parallel computing is very important to accelerate the performance of software systems. Additionally, considering that a recurring challenge is to process high data volumes continuously, \textit{stream processing} emerged as a paradigm and software architectural style. 
Several software systems rely on stream processing to deliver scalable performance, whereas open-source frameworks provide coding abstraction and high-level parallel computing. Although stream processing's performance is being extensively studied, the measurement of fault tolerance--a key abstraction offered by stream processing frameworks--has still not been adequately measured with comprehensive testbeds. In this work, we extend the previous fault recovery measurements with an exploratory analysis of the configuration space, additional experimental measurements, and analysis of improvement opportunities. We focus on robust deployment setups inspired by requirements for near real-time analytics of a large cloud observability platform. The results indicate significant potential for improving fault recovery and performance. However, these improvements entail grappling with configuration complexities, particularly in identifying and selecting the configurations to be fine-tuned and determining the appropriate values for them. Therefore, new abstractions for transparent configuration tuning are also needed for large-scale industry setups. We believe that more software engineering efforts are needed to provide insights into potential abstractions and how to achieve them. The stream processing community and industry practitioners could also benefit from more interactions with the high-level parallel programming community, whose expertise and insights on making parallel programming more productive and efficient could be extended.}


\maketitle

\section{Introduction}\label{sec1}

Stream processing embodies a computing paradigm and architectural pattern tailored for constructing distributed, event-driven software systems capable of handling immense volumes of data with minimal latency~\cite{Hirzel2013}. Contemporary stream processing paradigms empower software engineers to encapsulate their application's operational logic into abstract, high-level representations within directed acyclic graphs (DAGs). This level of abstraction eliminates the need for explicit definition of the physical execution plan and it is facilitated by mature, open-source stream processing frameworks like Spark~\cite{Armbrust2018}, Flink~\cite{Carbone2015}, and Kafka Streams~\cite{Wang2021}. These frameworks support developers in building highly scalable and efficient applications that process continuous data streams of massive volume. The frameworks provide high-level APIs and domain-specific languages to define operations such as filtering, transforming, aggregating, and merging data streams.

\subsection{Research Problem and Motivation}\label{subsec:problem}

Large industry companies frequently use and contribute to stream processing frameworks such as Flink, Kafka Streams, or Spark.
In contrast to more heavyweight standalone systems such as Flink or Spark, Kafka Streams is a lightweight programming library that can easily be integrated into existing applications~\cite{Bellemare2020}. 

Kafka Streams can provide simplicity and coding productivity to developers to write business logic for their stream processing applications without implementing and managing the underlying infrastructure. This is mostly due to Kafka Streams' high-level API that abstracts away much of the complexity of dealing with Kafka's lower-level APIs. Hence, Kafka Streams' is particularly interesting for use cases beyond pure analytics systems, for example, implementing core business logic in an event-driven microservice-based architecture~\cite{Katsifodimos2019,JSS2024}. %

Despite the great potential of Kafka Streams, the architecture that is tightly coupled with the Apache Kafka messaging system (or at least its protocols and APIs) implies a limited feature set compared to other systems. Moreover, Kafka Streams' functional (e.g., fault recovery) and non-functional requirements (e.g., performance) are still limited in comparison to its related approaches~\cite{vogel2024}.
Several works have been published that evaluate the performance of distributed stream processing frameworks~\cite{SEAA2023,JSS2024,ICPE2024} or optimizations~\cite{Hirzel2013,Herodotou2020,Vogel2022}.
Beyond measuring performance metrics such as throughput and latency, fault tolerance stands as a critical requirement for continuously operating production systems. Various levels of unpredictable failures can occur from the physical infrastructure to software layers~\cite{Wang2021}. Stream processing systems designed for continuous, low-latency processing demand swift recovery mechanisms to tolerate and mitigate failures effectively. 

Although fault tolerance guarantees are provided by modern stream processing frameworks~\cite{Fragkoulis2023}, it is notable that a relevant number of studies reported fault tolerance as challenging for stream processing~\cite{SEAA2023}. In our previous work~\cite{vogel2024},  we provided a comprehensive analysis of fault recovery performance, stability, and recovery time in a cloud-native environment with modern open-source frameworks (Flink, Kafka Streams, and Spark Structured Streaming). 

\begin{figure}[!h]
    \centering
    \includegraphics[width=\textwidth]{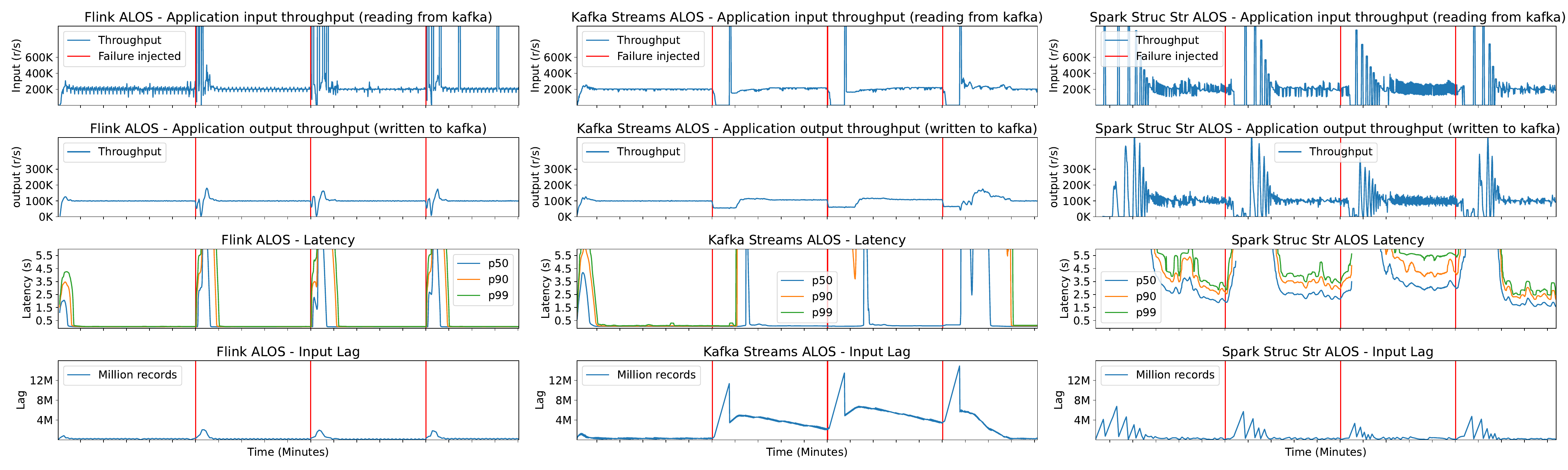}
    \caption{Two random pods executing worker instances are killed~\cite{vogel2024}.}
   \label{fig:characterization2podskill}
\end{figure}

A relevant observation from our previous work~\cite{vogel2024} is that Kafka Streams takes longer to recover from failures and presents volatile behavior. For instance,
\cref{fig:characterization2podskill} shows a scenario where periodically two random pods executing worker instances (entities executing the stream processing applications) are killed (see about the evaluation method and deployment setup in \cref{sec:setup}). After failures, Kafka Streams' partition assignment strategy, triggered by rebalances, causes its executions to accumulate more lag. This significantly increases event latency. Additionally, following failures, some worker instances may be assigned to more partitions or to partitions with accumulated records. Then, some workers become overloaded while others are underutilized. 

\textbf{Research Problem 1} tackled in this paper is how to improve Kafka Streams' fault recovery. 
During our investigation of the causes of the suboptimal fault recovery and rebalancing, we noticed a large configuration space that can be tuned and provide improvements.~\footnote{\href{https://kafka.apache.org/documentation/streams/developer-guide/config-streams.html}{https://kafka.apache.org/documentation/streams/developer-guide/config-streams.html}} However, it is usually complex for programmers and even system experts to understand and tune configurations of such large scale systems. Moreover, it is difficult to achieve configurations' generalization. Therefore, \textbf{Research~Problem~2} is related to understanding which configurations should be changed, to what values, and how this can be made more generic.

\subsection{Contributions}

This paper extends the fault recovery measurements with an exploratory analysis of the configuration space, experimental measurements, and analysis of improvement opportunities for robust deployment setups inspired by requirements for near real-time analytics of a cloud observability platform. We provide the following contributions: 

\begin{itemize}
    \item An extension of our experimental analysis~\cite{vogel2024}, providing a novel analysis of the experimental results, an exploratory analysis of the configuration space, and its impact on performance and fault recovery capabilities. 

    \item An extension of our method for automatic detection of failures and recovery~\cite{vogel2024}.

    \item Results discussion on fault recovery implications on industry deployments, configurations abstractions, and future research directions. %
    
\end{itemize}

\section{Background and Related Work}\label{sec:background}

This section introduces the fundamental concepts of modern stream processing frameworks, fault tolerance concepts, and overviews the state of the art.

\subsection{Distributed Stream Processing}\label{sec:background:stream-processing}

Stream processing frameworks perform operations such as filterings, transformations, or aggregations in near real-time on continuous streams of data~\cite{Hirzel2013}. 
State-of-the-art frameworks are designed for high throughput and low-latency processing, while also scaling with massive amounts of data~\cite{Fragkoulis2023,JSS2024}.
To address these requirements, they run in a distributed fashion on commodity hardware, nowadays often in managed cloud environments. A key advantage of stream processing frameworks is that they provide dataflow models that abstract aspects such as cluster management, state management, and time semantics from their users~\cite{Sax2018}. 
With such models, engineers describe the processing logic in DAGs of processing operators.
The frameworks allow the initiation of multiple worker instances across various compute nodes with multiple threads, each instance handling a distinct portion of the data.

While the isolated processing of data records remains unaffected by the assignment of data portions to worker instances, processing that depends on previous data records, such as aggregations, requires state management.
Similar to the MapReduce~\cite{Dean2008} programming model, keys are assigned to records before a stateful operation. This allows the stream processing frameworks to route all records with the same key to the same instance, where state synchronization among instances can be avoided.
When a processing operator modifies the key of a record and a subsequent operator performs a stateful operation, the framework divides the dataflow graph into subgraphs that can be independently processed by different worker instances. Popular stream processing frameworks include Apache Flink~\cite{Carbone2015}, Apache Kafka Streams~\cite{Wang2021}, and Apache Spark with its Structured Streaming engine~\cite{Armbrust2018}.

\subsection{Fault Tolerance}\label{sec:background-fault-tolerance}

Achieving consistency in stream processing has posed a longstanding research challenge, which is partly attributed to the lack of a formal problem specification~\cite{Wang2021}. Although executions can be subject to many failures, we expect computing systems to produce correct results despite the failures.
Such a requirement instigated research on fault tolerance, where notable advances achieved in distributed systems are applied to stream processing~\cite{Wang2021}.

In the absence of fault tolerance, stream processing applications would be forced to restart data processing from scratch whenever there is a loss due to a failure. Given that applications may have varying fault tolerance needs, the stream processing domain focuses on different processing semantics that precisely describe how a system is affected by failures~\cite{Fragkoulis2023}. The relevant terms to characterize the fault tolerance semantics guarantees are at-most-once semantics (AMOS), at-least-once semantics (ALOS), and exactly-once semantics (EOS). 
Currently, the majority of stream processing applications demand reliable executions with strong semantics guarantees. Thus, AMOS is obsolete~\cite{Fragkoulis2023} and we focus on ALOS and EOS. 

At-least-once semantics ensure that the system maintains consistent results even in the event of failures, preventing data loss. However, during recovery, duplicate output may occur as records might be processed more than once~\cite{Fragkoulis2023}. The frequency of such duplicates largely depends on the system's implementation, the type of failure encountered, and the timing of the failure during execution. Additionally, handling potential duplicates can be accomplished externally, for instance, through downstream systems filtering duplicates or database overwrites. With effective implementations, the fault tolerance guarantee provided by ALOS can be comparable to EOS.

Exactly-once semantics refers to the ability of the system to process the records exactly once without lost or duplicated results even under failures. Therefore, EOS requires additional implementation and has a performance impact on records processing~\cite{vanDongen2021a}. In this paper, we focus on semantics guarantees regarding the system output because this is the most relevant for the majority of use cases and also covers the guarantees in the internal state updates.
The stream processing frameworks provide different recovery semantic guarantees. This is important because the required guarantee varies according to the stream processing application and its use cases. In some scenarios, guaranteeing exactly-once can be of paramount importance, e.g., avoiding duplicate alerts in an observability platform. In other cases, at-least-once with rarely duplicated outputs is not equally detrimental, e.g., an information dashboard that receives twice the same CPU metric of a timestamp. Importantly, correct fault recovery is a crucial component of fault tolerance. While fault recovery focuses on restoring normal operation after a failure, fault tolerance aims to prevent downtime for high availability.

\subsection{Related Work}\label{sec:related}

Although many studies have presented various experimental evaluations of fault recovery in stream processing~\cite{SEAA2023}, a relevant one is~\cite{vanDongen2021a}. It dives into the implementation, performance, and efficiency of fault recovery in four state-of-the-art stream processing frameworks: Flink, Kafka Streams, Spark Streaming, and Spark Structured Streaming. In addition, they evaluated the behavior of these frameworks under different types of failures and settings, including master failure with and without high-availability setups, driver failures for Spark frameworks, worker failure with or without exactly-once semantics, and application failures.

\citet{Wang2021} explain Kafka Streams' processing logic and demonstrate how Kafka Streams behaves in practice with large-scale deployments and performance insights exhibiting its flexible and low-overhead trade-offs. \citet{Wang2022} propose a taxonomy of fault tolerance in stream processing. They also propose an evaluation framework tailored for fault tolerance, demonstrating experimental results on two representative open-source stream processing frameworks. 

The studies mentioned above fail to offer an updated and comprehensive analysis of the detrimental faults that may occur in production setups.
In this paper, we extend our previous work to tackle the research gaps discussed in~\cite{vogel2024}. We specifically concentrate on further analyzing the experimental results, exploring the configuration space for potential improvements, and assessing their impact on performance and fault recovery capabilities. \Cref{sec:setup} describes our experimental method.

\section{Experimental Method}\label{sec:setup}

This section describes our proposed evaluation method and setup.%

\subsection{Application Benchmark}\label{sec:benchmark}

\emph{ShuffleBench}~\cite{ICPE2024} is a benchmark for evaluating the performance of modern stream processing frameworks. It focuses on use cases where stream processing frameworks are mainly employed for \emph{shuffling} (i.e., re-distributing) data records to perform state-local aggregations, while the actual aggregation logic is considered as a black-box software component.
ShuffleBench is inspired by requirements for near real-time analytics of a large cloud observability platform, but it is highly configurable to allow domain-independent evaluations. It comes as a ready-to-use open-source software\footnote{\url{https://github.com/dynatrace-research/ShuffleBench}} utilizing existing Kubernetes tooling and providing implementations for different state-of-the-art stream processing frameworks.

\emph{ShuffleBench} adopts and extends the benchmarking framework Theodolite~\cite{EMSE2022} to automate the benchmark execution in Kubernetes-based cloud environments.
In short, data records are read from a messaging system (Kafka in our case), assigned to keys, and shuffled such that all records having the same key are forwarded to the same worker instance. There, a key-specific custom aggregation is performed and, optionally, an output event is created that is then written back to Kafka. We refer to our previous work for further details about \emph{ShuffleBench}~\cite{ICPE2024}.

\begin{figure}[!h]
    \centering
    \includegraphics[width=\linewidth]{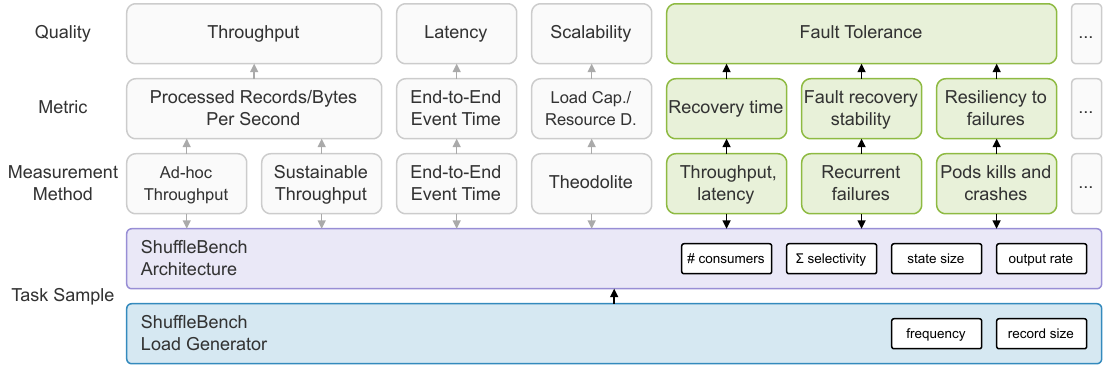}
    \caption{Overview of \emph{ShuffleBench} benchmark components~\cite{vogel2024}.}
    \label{fig:components}
\end{figure}

\emph{ShuffleBench} has a flexible architecture and supports several parametric configurations. In previous work, we extended \emph{ShuffleBench} for fault tolerance measurements~\cite{vogel2024}.
\Cref{fig:components} provides an overview of \emph{ShuffleBench's} benchmark components and their extension to fault tolerance, where the  metrics are described in \Cref{sec:metrics}.
We conduct our experimental evaluation in a Kubernetes cluster managed by the Elastic Kubernetes Service of Amazon Web Services with the following pods running in dedicated cloud instances: 2 load generators (\emph{m6i.xlarge} instances), 2 Kafka brokers (\emph{m6i.2xlarge}), a manager node used on Flink and Spark  (\emph{m6i.large}), and 8 pods for worker instances with 1 CPU, 3 GB of RAM (running on \emph{c5.large} instances). %

In our experiments, we set up 20\,000 real-time consumers that all have the same selectivity, which sums up to 50\,\%, meaning that each record is forwarded to $0.5$ consumers on average. Each consumer emits an output event for every record received. To consume and output results, 40 input and output Kafka partitions were used to maintain a suitable parallelism level that minimizes overheads. We run each experiment for 125~minutes in such a way that the experiments provide meaningful quantitative data, test with recurrent failures injected, and have sufficient time for every subsequent recovery. 

ShuffleBench runs on top of stream processing frameworks. In this work, we use Kafka Streams (version 3.5) as the stream processing framework. Kafka Streams supports ALOS by default, which is the fault recovery guarantee used in this work. 
Kafka Streams provides fault recovery with its strong integration with Kafka. Kafka acts as a messaging layer to safeguard state information. In-memory state stores are associated with a replicated changelog topic within Kafka, which tracks state updates. Kafka Streams applications are divided into multiple tasks, where each task is responsible for processing a subset of the input data partitions. Whenever a task interacts with the store, it maintains a changelog. If a task encounters failure, the system automatically restarts it on one of the remaining worker instances. The state stores can be restored by replaying their changelog, ensuring data consistency~\cite{Wang2021}. Kafka Streams' commit interval was set to 2 seconds. 

\subsection{Metrics} \label{sec:metrics}

We collected several metrics of the benchmarking application. The metric to measure processing speed is throughput, measured in records processed per time unit and collected every second as a moving average of the last 5 seconds. A higher throughput is usually desired in many use cases as it directly maps to the costs for operating the stream processing application in elastic cloud environments. We measure input throughput which refers to consuming records from the Kafka input topic, and output throughput which is the application output written into another Kafka topic. We also measure event latency (the lower latency tends to be better) as an average of a 10-second window. We show latency percentiles to measure the distribution of event times: the median (p50), 90\textsuperscript{th} percentile (p90), and 99\textsuperscript{th} percentile (p99). The latency measurements using the percentiles collected during the runtime are also complemented with the median latency from the entire execution. We use the median latency of the entire execution because this can provide a complementary view of latency during the entire execution.
We also measure the consumer lag corresponding to the number of records buffered in the input topic.

The fault recovery on stream processing is measured with the metrics of fault recovery performance, and the median fault recovery time. We also show the CPU utilization from the cloud instances that run the worker pods (collected every 2 seconds). To enhance visual clarity, the CPU utilization is displayed using a 10-value moving window average.

\subsection{Failure Injection Scenarios} \label{sec:scenarios}

\begin{figure}[!h]
    \centering
    \includegraphics[width=.5\textwidth]{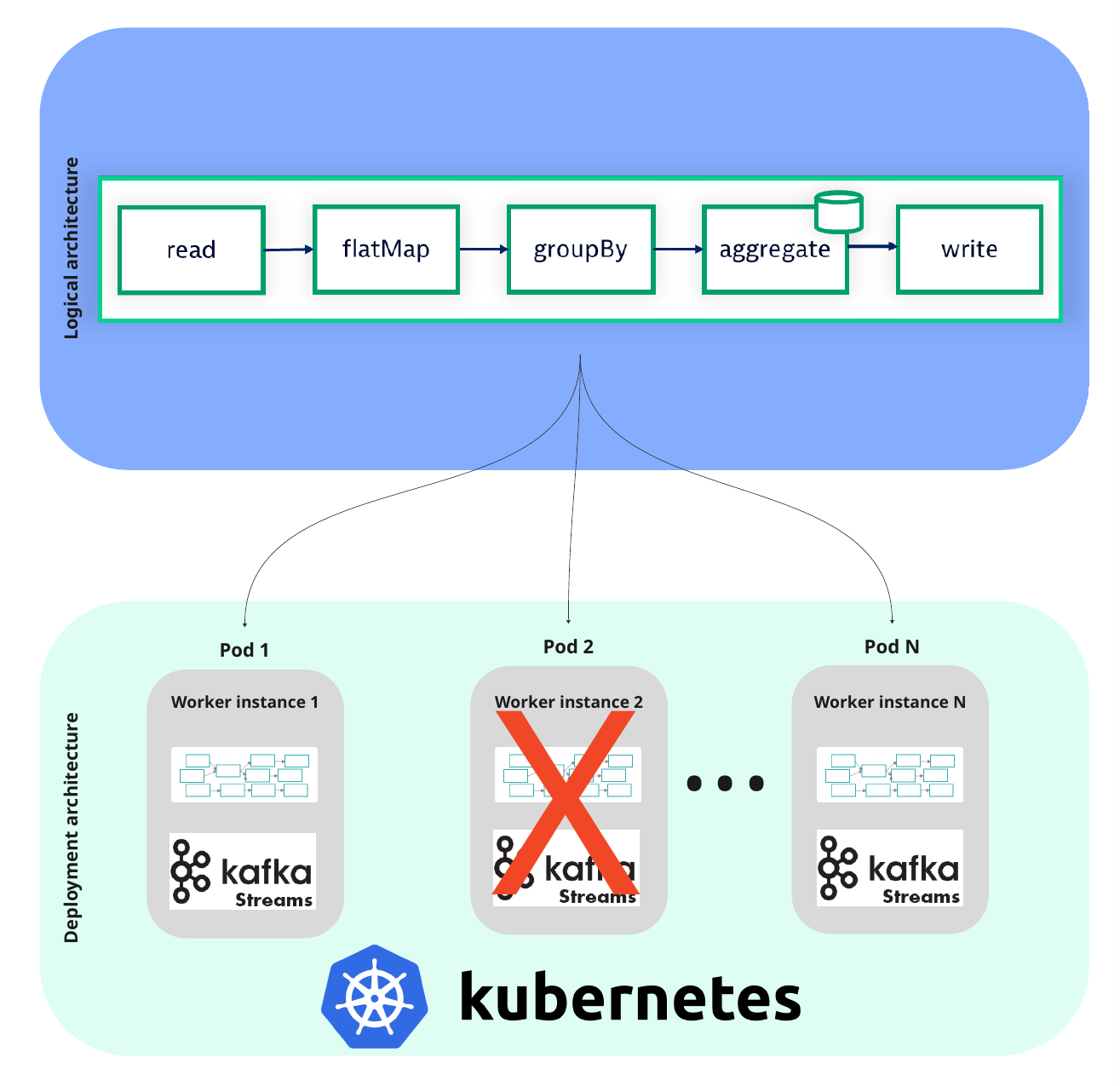}
    \caption{High-level view of \emph{ShuffleBench's} execution with a pod failure.}
    \label{fig:failure-example}
\end{figure}

\Cref{fig:failure-example} illustrates the used architecture and the failure scenario. The application benchmark (see \cref{sec:benchmark}) runs on a microservice-based architecture provided by Kubernetes pods, where the pods are executed on top of virtual machines. The application benchmark logical architecture is executed on worker pod instances,\footnote{The actual mapping varies according to the stream processing framework execution model and scheduling.} which in this example with Kafka Streams as the stream processing framework. In perfect execution conditions, the application executes without any disturbance from the infrastructure. However, in reality, many entities can fail~\cite{Wang2021}. \textbf{Pods kill} is the scenario considered in this paper, where random pods executing worker instances are deleted, which is illustrated in \cref{fig:failure-example} when a pod is killed at a given moment of the execution.

It is important to note that when the pods are no longer accessible, both the records they were processing and their associated state are lost. Therefore, the frameworks have to trigger recovery from such a failure. Deployed with Kubernetes, new pods will be created and added to the pool of worker instances within an interval of seconds (usually taking from 2 to 10 seconds). We experimented with killing 1, 2, and 4 worker pods at once. Such values were determined to simulate simple and critical failure scenarios. Such failures are recurrently injected every 12 minutes to evaluate the fault recovery stability. This interval was defined for a reasonable recovery time between failures and also to avoid unnecessary long executions that would waste resources. 

This work uses experimental scenarios inspired by chaos engineering~\cite{Basiri2016}. Failures are injected using Chaos Mesh,\footnote{\url{https://github.com/chaos-mesh/chaos-mesh}} an open-source chaos engineering platform integrated with the Kubernetes deployment. 

\subsection{Configuration Space Explored}\label{sec:configurations}

Considering the research problems tackled in this work (see~\cref{subsec:problem}) and the large configuration space available,\footnote{\href{https://kafka.apache.org/documentation/streams/developer-guide/config-streams.html}{https://kafka.apache.org/documentation/streams/developer-guide/config-streams.html}} we are interested in exploring the potential of tuning configurations to improve the recovery speed, performance after failures, and avoid the demands of additional computing resources. From Kafka Streams' community, one of the configurations mostly tuned in production is adding standby replicas.\footnote{The aim is to establish a predetermined number of replicas per store and maintain their synchronization, contingent upon a sufficient number of operational worker instances. In the event of task recovery within worker instances, priority is assigned to those equipped with standby replicas tailored to the specific task.} In practice, standby replicas demand much more resources. Therefore, we are interested in other configurations that can still improve fault recovery without demanding more resources, which would still be a comparable approach to other stream processing frameworks.

Investigating the causes of the performance overhead in Kafka Streams' rebalance protocol, we noticed that one of the main reasons is that the convergence is too slow after failures. The failure of worker instances triggers a rebalance, where then the partitions affected by failure are usually unevenly redistributed to the remaining instances. After a redistribution, follow-up rebalances are needed to achieve an even partition distribution to improve the load balance.\footnote{The number of rebalances necessary to achieve even partitions' distribution can vary significantly.} However, the first Kafka Streams' rebalance is only triggered 10 minutes after the failure with the default configurations. In our exploratory research and interactions with the community experts, we noted that this default period can be changed with the \textit{probing.rebalance.interval.ms} configuration, which represents the maximum waiting period before initiating a rebalance, aimed at verifying the readiness of warm-up replicas for them to be deemed sufficiently up to date to tasks reassignment. When a given task needs to be moved to another worker instance, a warm-up replica is placed in the destination worker instance before it receives a task reassignment. Hence, the worker instance can ``warm-up'' the state in the background without interrupting the active processing, having a smooth transition. 

In this work, we extend the fault recovery measurements with an alternative Kafka Streams approach with configuration tuning. Specifically, we evaluate a shorter rebalance interval set to a minimum of 1 minute, and we allow more warm-up replicas up to a value equal to the parallelism level (8) used in our experiments. These configurations can in theory speed up the fault recovery by being faster (shorter rebalance interval) and more robust rebalances (more warm-up replicas enable more tasks to be migrated per rebalance), which can enable an optimal number of partitions distribution to the worker instances available. An optimal distribution can provide a true fault recovery with load balance and improve the application performance. We compare \textit{Kafka Streams with tuning} (abbreviated \textit{Kafka Streams w/ tuning}) to Kafka Streams using default executions (referred to as \textit{Kafka Streams default}).

Although many other configurations and values could be evaluated, the main goal of this work is to measure the potential gains of these configurations in a robust benchmarking setup. Moreover, we discuss in \cref{sec:discussion} what can be improved in the current and future approaches.

\section{Fault Recovery Performance}\label{sec:characterization}

This section intends to demonstrate the potential impacts of faults and their recovery on the different approaches, \textit{Kafka Streams default} and \textit{Kafka Streams with tuning}.

\begin{figure}[!h]
	\centering
	\includegraphics[width=\textwidth]{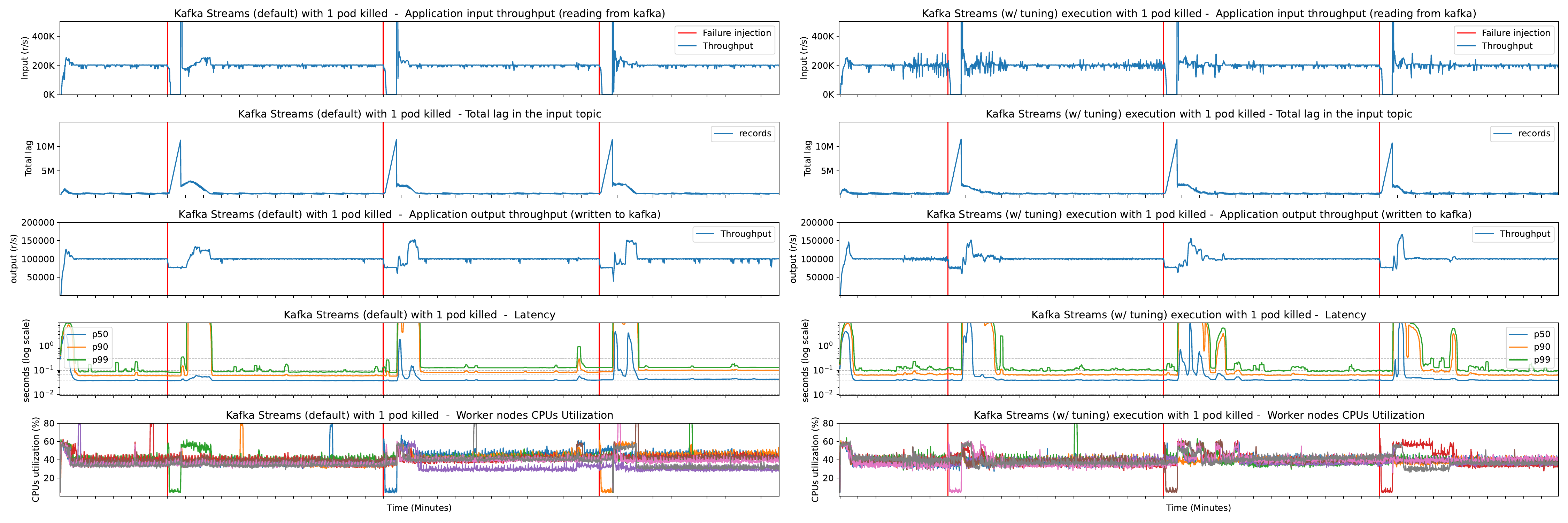} 
	\caption{Fault recovery characterization: Random pods executing worker instances are killed. \textit{Default Kafka Streams} (left side) and \textit{Kafka Streams w/ tuning} (right side).}
	\label{fig:1pod-kill}
\end{figure}

\Cref{fig:1pod-kill} shows how the executions behave when pods executing worker instances are killed (represented by the red line). To have more comprehensive characterizations, we present results with three failures that occur periodically every 12 minutes. It is also important to note that the plots show different metrics (see y-axis labels) over time. When a failure occurs, the worker instance impacted is lost. Then, the Kubernetes infrastructure creates a new pod that is added to the pool of worker instances. Moreover, the stream processing frameworks are expected to start scheduling computations to new worker instances. This is done in Kafka Streams by rebalancing and assigning partitions to new worker instances. 

Comparing the approaches, it is notable in \cref{fig:1pod-kill} that the default execution achieves a stabler execution in terms of throughput and latency. Moreover, the latency seemingly stabilizes quicker in the default approach after the failures. However, the p90 and p99 latencies do not normalize after the second failure in Kafka Streams default execution. For instance, the average latency p90 stabilized after the second failure is 83.99 milliseconds (ms), while the reference one before the first failure is 60.81 ms (32.03\% lower). This is due to the known limitation in the rebalancing protocol (see \cref{subsec:problem}). Another aspect of such a limitation can be noted in imbalances in the worker nodes' CPUs utilization after the second failure.

In contrast, the execution with configuration tuning returns to latencies comparable to the ones before the failure. This is due to the earlier rebalances and optimal partition distribution provided with the tuned configurations (shorter rebalance interval and faster task migration). Moreover, evidence supporting the improvements is the fact that after the failures the \textit{Kafka Streams w/ tuning} execution achieves a similar CPUs utilization in the worker nodes, which indicates a balanced assignment of partitions and tasks to worker instances.

\begin{figure}[!h]
	\centering
	\includegraphics[width=\textwidth]{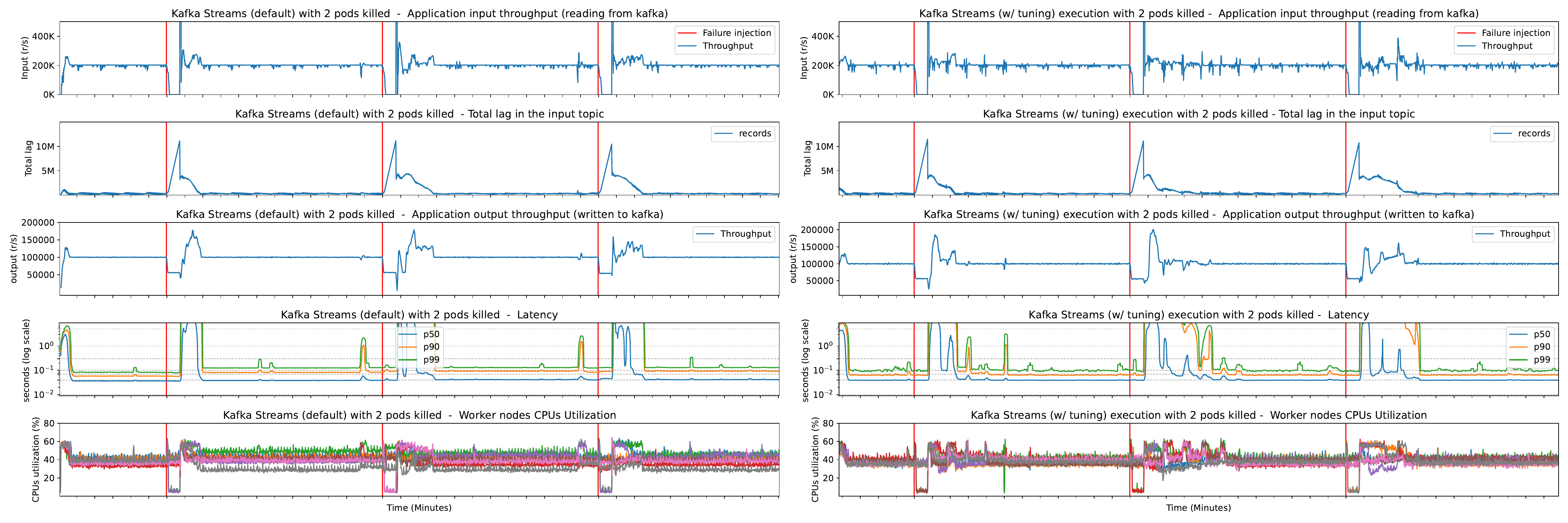} 
	\caption{Fault recovery characterization: Two random pods executing worker instances are killed. \textit{Default Kafka Streams} (left side) and \textit{Kafka Streams w/ tuning} (right side).}
	\label{fig:2pod-kill}
\end{figure}

\Cref{fig:2pod-kill} shows the scenario where two random pods executing worker instances are killed. We expect this scenario to be more challenging as more workers failing means that more state can be lost and more records need to be replayed from the input source and reprocessed. Compared to \cref{fig:1pod-kill}, \cref{fig:2pod-kill} shows similar performance, fault recovery, and resource consumption observations. However, it is noticeable in \cref{fig:2pod-kill} that Kafka Streams default's execution already faces significant imbalances after the first failure. On the other hand, the \textit{Kafka Streams w/ tuning} execution recovers after failures to a state comparable to the reference. However, the correct recovery comes at the expense of slight fluctuations in the first minutes after failures, which is due to the rebalances necessary to achieve an optimal distribution of partitions and tasks.

\begin{figure}[!h]
	\centering
	\includegraphics[width=\textwidth]{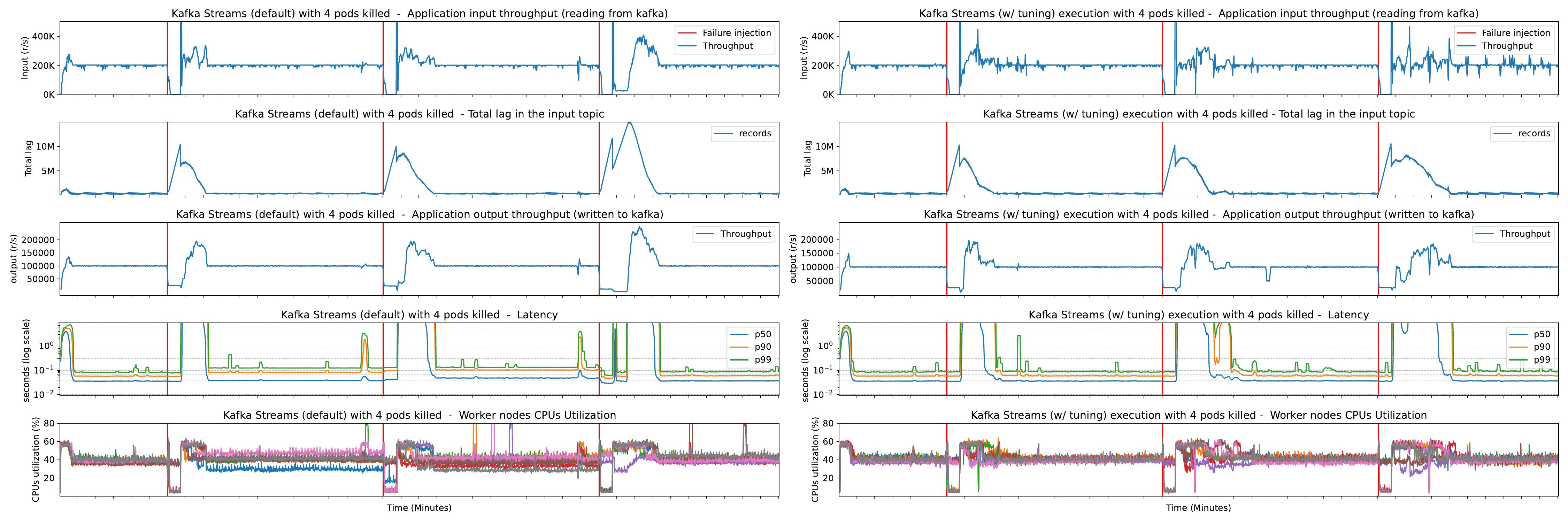} 
	\caption{Fault recovery characterization: Four random pods executing worker instances are killed. \textit{Default Kafka Streams} (left side) and \textit{Kafka Streams w/ tuning} (right side).}
	\label{fig:4pod-kill}
\end{figure}

\Cref{fig:4pod-kill} shows a scenario of more detrimental failures where four worker pods are killed at the same time. Generally, the results from \cref{fig:4pod-kill} complement and further support the observations from the scenarios of one and two pods killed. It is also notable in \cref{fig:4pod-kill} that the Kafka Streams execution with default configurations accumulates more lag after failures. Moreover, in terms of output throughput after the failures, the default execution has a lower throughput and takes longer to normalize.

\begin{figure}[!h]
	\centering
	\includegraphics[width=.75\textwidth]{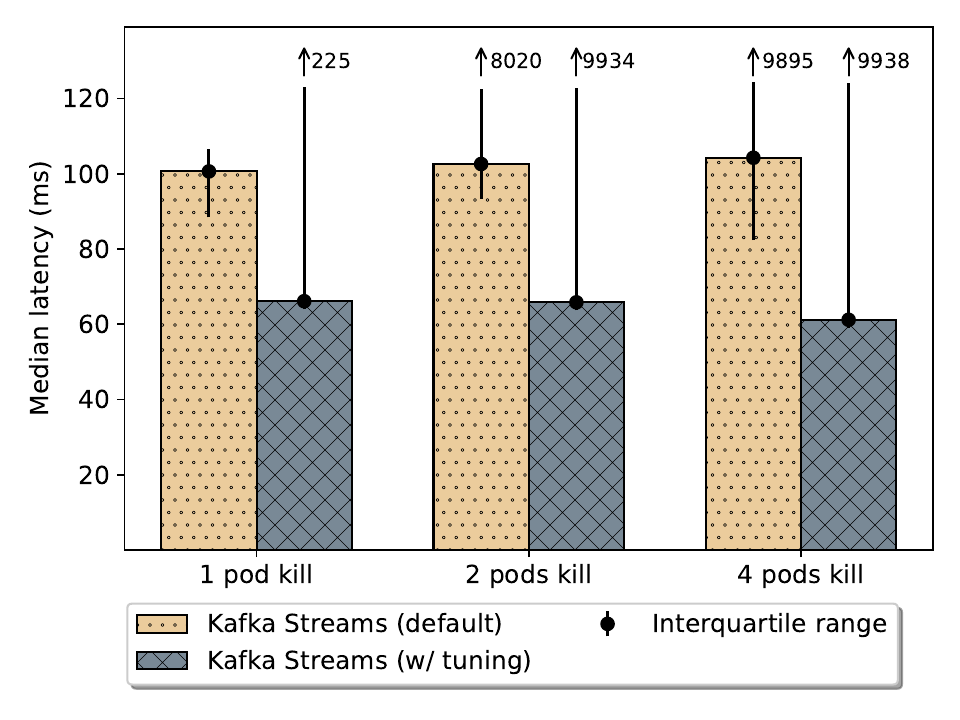} 
	\caption{Median latency of the p90 percentile.}
	\label{fig:average-latency}
\end{figure}

In addition to characterizing the approaches' fault recovery capabilities, insights from the quality of services (QoS) metrics are also relevant. Considering the relevance of low latency for stream processing applications, \cref{fig:average-latency} shows the median latency of the representative p90 percentile from the entire executions, where such executions were characterized in \cref{fig:1pod-kill,fig:2pod-kill,fig:4pod-kill}. The median latency is shown as it can capture the stable latency values after the failure recovery, where we expect such a period to be higher than the failure recovery. Therefore, the median latency can provide relevant values by leaving out the latency measurements recorded during fault recovery.

The results from \cref{fig:average-latency} demonstrate that the median latency can be stable across different failures. Moreover, \textit{Kafka Streams w/ tuning} achieves a significantly lower (better) median latency than the default approach with contrasts ranging from 41.38\% to 52.07\%. The executions of \textit{Kafka Streams w/ tuning} show higher values in the upper interquartile range (75\% upper quantile), which is due to the slightly longer recoveries with very high latencies due to the many rebalances necessary.

\section{Fault Recovery Time}\label{sec:recovery-time}

The precise fault recovery time is another relevant metric. Considering that there is a lack of approaches and tools to automate the measurements of fault recovery times across many failures, we show a proposed detector in previous work~\cite{vogel2024}. The detector considers the moving average and standard deviation to automatically detect the failure as well as the time to recovery. The detector extracts the expected average and standard deviation for both throughput and latency based on the regular processing behavior that follows the warm-up period. After the execution warm-up, there is a reference time frame before the first failure that is used as stable behavior. 

In short, the detector marks as the point of failure injection the cases where the moving average and standard deviation exceed a parameterized threshold value. Afterward, the other points of failure are strictly marked every 12 minutes, which is also a parametric value defined according to our experimental method. 

\begin{figure}
    \centering
    \includegraphics[width=0.9\textwidth]{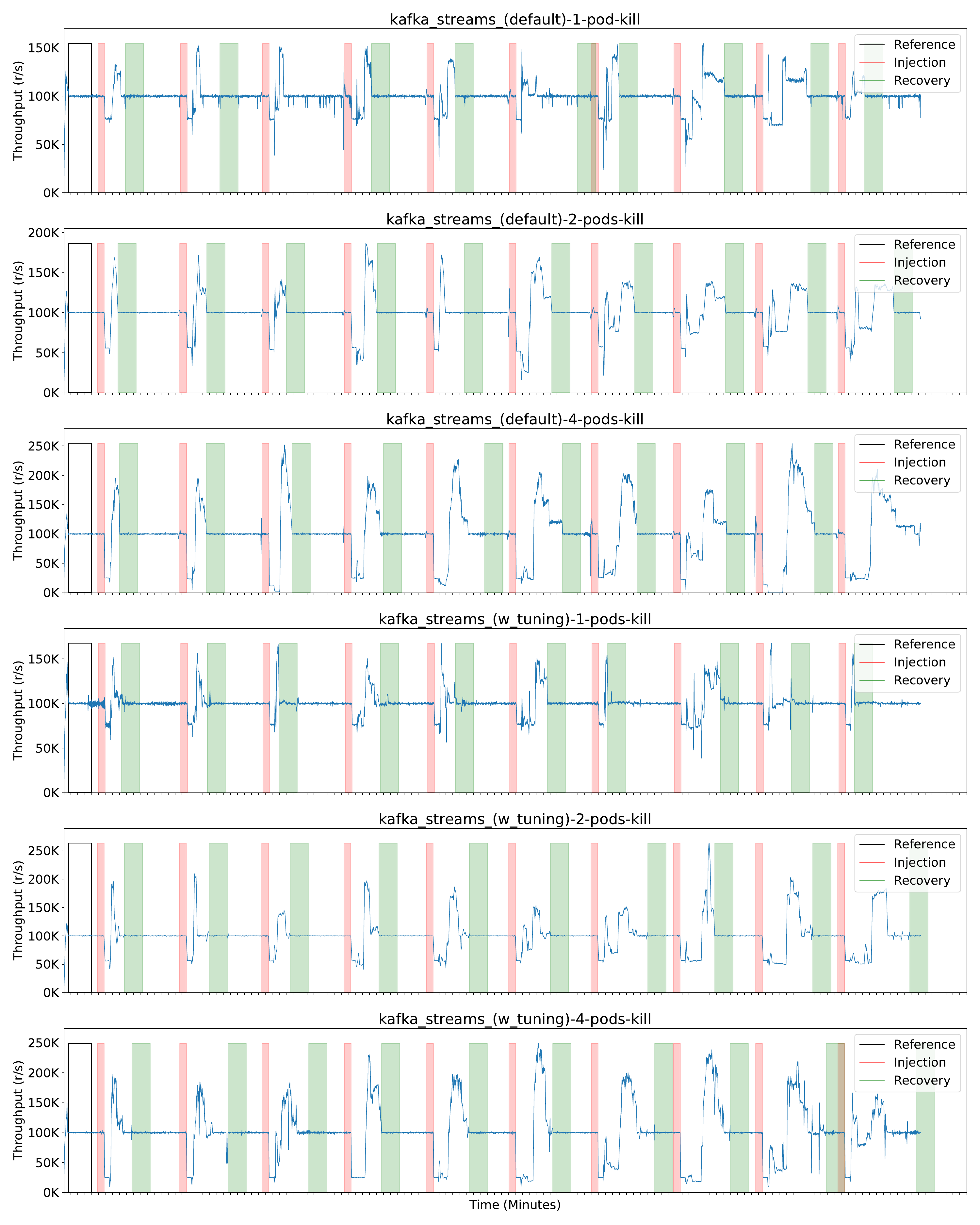}
    \caption{Characterization of our approach for automating measurements of fault recovery times. Stable behavior is depicted as square shapes that are references for training. Failures are highlighted in red, while recovery detections are highlighted in green.}
    \label{fig:recovery-approach-characterization}
\end{figure}

Similarly, the recovery time was detected when the estimators returned within the same threshold (15\%) for a larger time frame (at least  160 seconds). We expect throughput and latency to stabilize after the recovery time, which should be only detected when the execution returns to a normal state. It is important to note that our approach for detection failures and their recovery was configured to automate our measurements. Considering that it is parametric and customizable, we expect it to be also applicable to wider scenarios.\footnote{We plan to share the tooling and all the parameters used as supplementary material.} Moreover, we believe that there are no ground truths about these parameter values used. In fact, we focus more on having a formal parametric method to automate the measurements that is less arbitrary and avoids biased analysis. The values detected can be an estimation of the recovery time that we expect to be ascertained with visual inspections in the metrics traces. 

\Cref{fig:recovery-approach-characterization} exemplifies the trace of automatically detected failures and their recoveries. For conciseness, we show here only the throughput case. It is notable in \cref{fig:recovery-approach-characterization} that the recovery time can vary between recurrent failures, which is mostly due to what worker instances are affected, their state w.r.t. the last saved checkpoint, and the state size to be restored.

\begin{figure}[htbp]
  \centering
  \begin{subfigure}[htbp]{.49\textwidth} %
    \centering
    \includegraphics[width=1\textwidth]{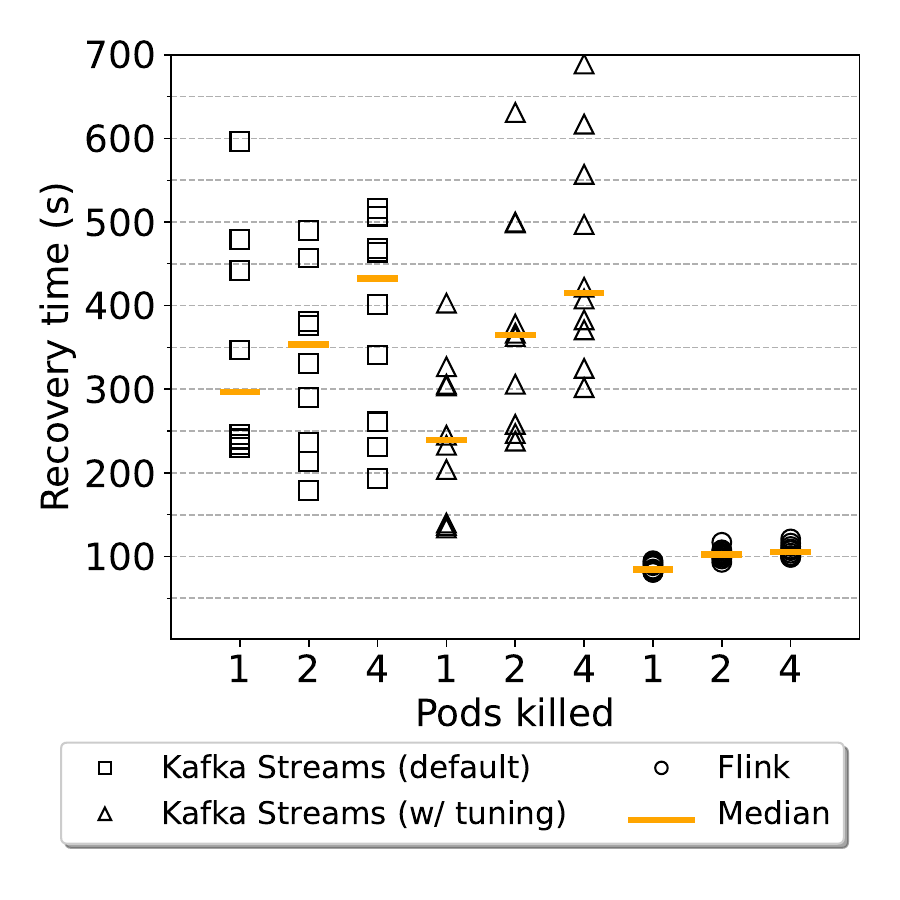}
    \caption{Recovery time of the throughput.}
    \label{fig:recovery-throughput}
  \end{subfigure}
  \begin{subfigure}[htbp]{.49\textwidth} %
    \centering
    \includegraphics[width=1\textwidth]{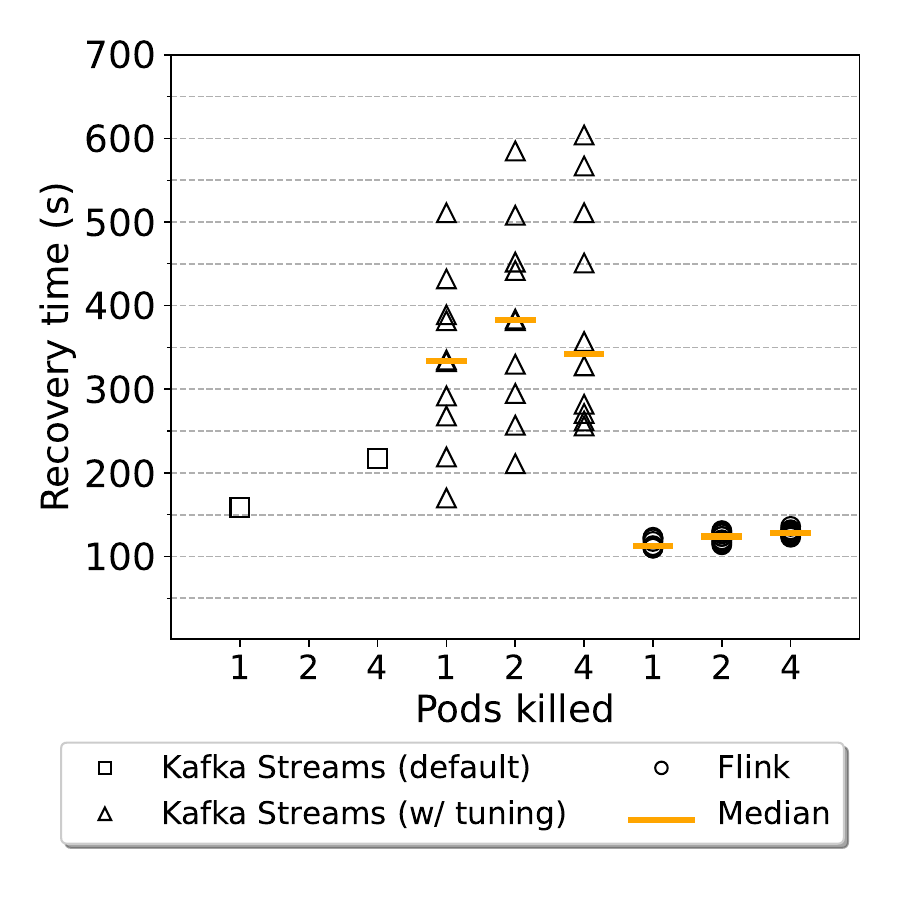}
    \caption{Recovery time of the latency p90.}
    \label{fig:recovery-latency}
  \end{subfigure}
  \caption{Recovery times identified by our approach across 10 measurements per framework/configurations and pods killed scenarios.}%
  \label{fig:recovery-times}
\end{figure}

Our approach to automating fault recovery times allows one to conduct relevant measurements and analysis. \Cref{fig:recovery-times} summarizes the results of the fault recovery times exemplified in \cref{fig:recovery-approach-characterization} with output throughput and latency (p90) as the metrics. Moreover, \cref{fig:recovery-times} shows Flink's (version 1.17) recovery time from our previous work~\cite{vogel2024}, where Flink can be seen as the state of the art approach to be used as a baseline. Regarding the throughput recovery depicted in \cref{fig:recovery-throughput}, it is notable that Flink's executions still achieve the fastest recovery in terms of throughput. Moreover, Kafka Streams' executions with tuning can provide slight improvements in the recovery of the application's output throughput, which is relevant to reducing downtimes.  

The latency recovery is depicted in~\cref{fig:recovery-latency} where Flink again achieved the fastest recovery. In contrast, Kafka Streams' default executions only fully recovered in the 12-minute period (before the next subsequent failure was injected) in two cases out of 30 failures injected. This outcome is mostly due to the slow rebalancing after failures causing limited latency recovery (discussed in \cref{sec:characterization}). It is important to note that the decision of whether a given execution instance is recovered or not is made systematically by our automated approach. Moreover, the interpretation of a complete fault recovery can vary in other use cases or QoS requirements. 

\Cref{fig:recovery-latency} shows that Kafka Streams' executions with tuning recovered from failures, achieving median recovery times between 300 and 400 seconds. Finally, the latency recovery achieved on the execution with tuning evinces Kafka Streams' potential for configuration tuning. However, this comes at the expense of significant complexity to understand which configurations should be tuned and with what values.

\section{Discussion}\label{sec:discussion}

The performance gains achieved with the \textit{Kafka Streams w/ tuning} approach is of paramount importance. This is mostly because the fault recovery capabilities impact the quality of service and are a criterion for choosing stream processing frameworks. However, achieving such gains is conditioned to significant complexities in understanding which configurations should be tuned and with what values, which can be a burden for developers. Moreover, these configurations tend to be limited in terms of generalizability to other frameworks and deployments.

Frameworks such as Spark and Flink~\cite{guo2021} received improvements in configuration tuning and abstractions for developers~\cite{li2023,guo2021}. Automatic parameter and configuration tuning is possible~\cite{Herodotou2020}. However, it is debatable whether the research community should continue researching approaches to tune only a unique framework. Generalizability is still a challenge with limited use cases of improvements.

Another potential alternative is prompting large language models (LLMs) to automatically tune configurations~\cite{zhang2024}. However, we noticed that currently GPT-3.5 and industry-grade copilots such as Microsoft's are still unable to identify the relevant configurations to be tuned in Kafka Streams. Our understanding is that many further developments are needed before LLMs are integrated and empirically evaluated for our concrete scenario. Training such models without previous data but using run-time data is another challenge that demands future studies.

Focusing on high-level abstractions, run-time self-adaptation can also be a potential approach to release programmers from the burden of finding optimal configuration values. One direction is to expand what was achieved with dynamic checkpointing interval optimizations~\cite{geldenhuys2022}. However, run-time self-adaptation is still limited in terms of which configurations can be dynamically changed at run-time~\cite{Vogel2022}.  However, we believe that more software engineering efforts are needed to provide insights into potential abstractions and how to achieve them. The research problems tackled in our work would also certainly benefit from the intersection with the high-level parallel programming community, whose expertise and insights on making parallel programming more productive and efficient could be extended to additional scenarios.

\section{Conclusions}\label{sec:conclusions}

Ensuring fault tolerance in data-intensive, event-driven applications is crucial for successful industry deployments. In this paper, we introduced an enhanced evaluation methodology alongside fault recovery analysis to further advance this critical aspect. Such an analysis was enabled by \emph{ShuffleBench's} flexible deployment~\cite{ICPE2024} and extending from our previous work~\cite{vogel2024} the fault recovery analysis. 

Regarding the research problems, \textbf{Research~Problem~1} on improving Kafka Streams' fault recovery was tackled as our results demonstrate significant fault recovery and performance gains. However, achieving such gains comes at the cost of complexities regarding what configurations should be tuned and with what values. Our perspective is that new or improved abstractions are also needed for transparent configuration tuning targeting and measuring large-scale industry setups. Therefore, \textbf{Research~Problem~2} necessitates further research and novel approaches to enhance the generalizability of configuration tuning and provide abstractions for developers.

The insights from our experimental results and exploitation of potential fault recovery improvements have also a direct impact on advanced optimizations necessary in modern stream processing, such as run-time self-adaptations for auto-scaling. This connection arises from the fact that implementing run-time adaptations in distributed executions inherently involves fault recovery procedures. Furthermore, we believe that our experimental methodology possesses a level of generality that extends its utility beyond our immediate focus. Specifically, it can be applied to evaluating other domains demanding fault recovery in distributed systems deployed as microservices (e.g., in Kubernetes). This flexibility is primarily from the comprehensive nature of our methodology, which includes insights into the use of chaos engineering principles in long-running and distributed computing systems.

We expect the threats to the validity discussed in our previous works~\cite{vogel2024,ICPE2024} to be also applicable here. Relevant configurations such as the failure injection on worker instances and the number of affected workers were defined based on insights from industry deployments. In such deployments, the failure probability of worker instances increases due to the deployment of hundreds or thousands of them.

We believe that future analyses are still needed.
Our perspective is that Kafka Streams can be an appealing approach for scalability in JVM-based applications. While employing Kafka Streams' standby replicas necessitates additional resources, in the future, we intend to assess this deployment in terms of the trade-offs between the advantages of fault recovery and the additional resource costs. Moreover, from a high-level parallel computing perspective, we believe that complementary studies are needed also with software engineering methods to measure coding productivity with high-level frameworks for scalable stream processing, complementary to what is being done with parallel programming interfaces~\cite{andrade2021}. We also intend to extend our work to measure fault recovery for scenarios where exactly-once recovery guarantees are needed.

\bmhead{Acknowledgements}

We would like to thank the Johannes Kepler University Linz and Dynatrace for co-funding this research.

\bibliography{preprint}%


\begin{thebibliography}{24}
\ifx \bisbn   \undefined \def \bisbn  #1{ISBN #1}\fi
\ifx \binits  \undefined \def \binits#1{#1}\fi
\ifx \bauthor  \undefined \def \bauthor#1{#1}\fi
\ifx \batitle  \undefined \def \batitle#1{#1}\fi
\ifx \bjtitle  \undefined \def \bjtitle#1{#1}\fi
\ifx \bvolume  \undefined \def \bvolume#1{\textbf{#1}}\fi
\ifx \byear  \undefined \def \byear#1{#1}\fi
\ifx \bissue  \undefined \def \bissue#1{#1}\fi
\ifx \bfpage  \undefined \def \bfpage#1{#1}\fi
\ifx \blpage  \undefined \def \blpage #1{#1}\fi
\ifx \burl  \undefined \def \burl#1{\textsf{#1}}\fi
\ifx \doiurl  \undefined \def \doiurl#1{\url{https://doi.org/#1}}\fi
\ifx \betal  \undefined \def \betal{\textit{et al.}}\fi
\ifx \binstitute  \undefined \def \binstitute#1{#1}\fi
\ifx \binstitutionaled  \undefined \def \binstitutionaled#1{#1}\fi
\ifx \bctitle  \undefined \def \bctitle#1{#1}\fi
\ifx \beditor  \undefined \def \beditor#1{#1}\fi
\ifx \bpublisher  \undefined \def \bpublisher#1{#1}\fi
\ifx \bbtitle  \undefined \def \bbtitle#1{#1}\fi
\ifx \bedition  \undefined \def \bedition#1{#1}\fi
\ifx \bseriesno  \undefined \def \bseriesno#1{#1}\fi
\ifx \blocation  \undefined \def \blocation#1{#1}\fi
\ifx \bsertitle  \undefined \def \bsertitle#1{#1}\fi
\ifx \bsnm \undefined \def \bsnm#1{#1}\fi
\ifx \bsuffix \undefined \def \bsuffix#1{#1}\fi
\ifx \bparticle \undefined \def \bparticle#1{#1}\fi
\ifx \barticle \undefined \def \barticle#1{#1}\fi
\bibcommenthead
\ifx \bconfdate \undefined \def \bconfdate #1{#1}\fi
\ifx \botherref \undefined \def \botherref #1{#1}\fi
\ifx \url \undefined \def \url#1{\textsf{#1}}\fi
\ifx \bchapter \undefined \def \bchapter#1{#1}\fi
\ifx \bbook \undefined \def \bbook#1{#1}\fi
\ifx \bcomment \undefined \def \bcomment#1{#1}\fi
\ifx \oauthor \undefined \def \oauthor#1{#1}\fi
\ifx \citeauthoryear \undefined \def \citeauthoryear#1{#1}\fi
\ifx \endbibitem  \undefined \def \endbibitem {}\fi
\ifx \bconflocation  \undefined \def \bconflocation#1{#1}\fi
\ifx \arxivurl  \undefined \def \arxivurl#1{\textsf{#1}}\fi
\csname PreBibitemsHook\endcsname

\bibitem[\protect\citeauthoryear{Hirzel et~al.}{2014}]{Hirzel2013}
\begin{botherref}
\oauthor{\bsnm{Hirzel}, \binits{M.}},
\oauthor{\bsnm{Soul\'{e}}, \binits{R.}},
\oauthor{\bsnm{Schneider}, \binits{S.}},
\oauthor{\bsnm{Gedik}, \binits{B.}},
\oauthor{\bsnm{Grimm}, \binits{R.}}:
A catalog of stream processing optimizations.
ACM Comput Surv
\textbf{46}(4)
(2014)
\end{botherref}
\endbibitem

\bibitem[\protect\citeauthoryear{Armbrust et~al.}{2018}]{Armbrust2018}
\begin{bchapter}
\bauthor{\bsnm{Armbrust}, \binits{M.}},
\bauthor{\bsnm{Das}, \binits{T.}},
\bauthor{\bsnm{Torres}, \binits{J.}},
\bauthor{\bsnm{Yavuz}, \binits{B.}},
\bauthor{\bsnm{Zhu}, \binits{S.}},
\bauthor{\bsnm{Xin}, \binits{R.}},
\bauthor{\bsnm{Ghodsi}, \binits{A.}},
\bauthor{\bsnm{Stoica}, \binits{I.}},
\bauthor{\bsnm{Zaharia}, \binits{M.}}:
\bctitle{{Structured Streaming}: A declarative api for real-time applications in {Apache Spark}}.
In: \bbtitle{Proceedings of the 2018 International Conference on Management of Data},
pp. \bfpage{601}--\blpage{613}.
\bpublisher{Association for Computing Machinery},
\blocation{New York, NY, USA}
(\byear{2018})
\end{bchapter}
\endbibitem

\bibitem[\protect\citeauthoryear{Carbone et~al.}{2015}]{Carbone2015}
\begin{botherref}
\oauthor{\bsnm{Carbone}, \binits{P.}},
\oauthor{\bsnm{Katsifodimos}, \binits{A.}},
\oauthor{\bsnm{Ewen}, \binits{S.}},
\oauthor{\bsnm{Markl}, \binits{V.}},
\oauthor{\bsnm{Haridi}, \binits{S.}},
\oauthor{\bsnm{Tzoumas}, \binits{K.}}:
{Apache Flink}: Stream and batch processing in a single engine.
Bulletin of the IEEE Computer Society Technical Committee on Data Engineering
\textbf{36}(4)
(2015)
\end{botherref}
\endbibitem

\bibitem[\protect\citeauthoryear{Wang et~al.}{2021}]{Wang2021}
\begin{bchapter}
\bauthor{\bsnm{Wang}, \binits{G.}},
\bauthor{\bsnm{Chen}, \binits{L.}},
\bauthor{\bsnm{Dikshit}, \binits{A.}},
\bauthor{\bsnm{Gustafson}, \binits{J.}},
\bauthor{\bsnm{Chen}, \binits{B.}},
\bauthor{\bsnm{Sax}, \binits{M.J.}},
\bauthor{\bsnm{Roesler}, \binits{J.}},
\bauthor{\bsnm{Blee-Goldman}, \binits{S.}},
\bauthor{\bsnm{Cadonna}, \binits{B.}},
\bauthor{\bsnm{Mehta}, \binits{A.}},
\bauthor{\bsnm{Madan}, \binits{V.}},
\bauthor{\bsnm{Rao}, \binits{J.}}:
\bctitle{Consistency and completeness: Rethinking distributed stream processing in {Apache Kafka}}.
In: \bbtitle{SIGMOD/PODS '21},
pp. \bfpage{2602}--\blpage{2613}.
\bpublisher{Association for Computing Machinery},
\blocation{New York, NY, USA}
(\byear{2021})
\end{bchapter}
\endbibitem

\bibitem[\protect\citeauthoryear{Bellemare}{2020}]{Bellemare2020}
\begin{bbook}
\bauthor{\bsnm{Bellemare}, \binits{A.}}:
\bbtitle{Building Event-Driven Microservices}.
\bpublisher{O'Reilly Media, Inc},
\blocation{US}
(\byear{2020})
\end{bbook}
\endbibitem

\bibitem[\protect\citeauthoryear{Katsifodimos and Fragkoulis}{2019}]{Katsifodimos2019}
\begin{bchapter}
\bauthor{\bsnm{Katsifodimos}, \binits{A.}},
\bauthor{\bsnm{Fragkoulis}, \binits{M.}}:
\bctitle{Operational stream processing: Towards scalable and consistent event-driven applications}.
In: \bbtitle{Advances in Database Technology - 22nd International Conference on Extending Database Technology},
pp. \bfpage{682}--\blpage{685}.
\bpublisher{OpenProceedings.org},
\blocation{Lisbon}
(\byear{2019})
\end{bchapter}
\endbibitem

\bibitem[\protect\citeauthoryear{Henning and Hasselbring}{2024}]{JSS2024}
\begin{barticle}
\bauthor{\bsnm{Henning}, \binits{S.}},
\bauthor{\bsnm{Hasselbring}, \binits{W.}}:
\batitle{Benchmarking scalability of stream processing frameworks deployed as microservices in the cloud}.
\bjtitle{Journal of Systems and Software}
\bvolume{208},
\bfpage{111879}
(\byear{2024})
\end{barticle}
\endbibitem

\bibitem[\protect\citeauthoryear{Vogel et~al.}{2024}]{vogel2024}
\begin{botherref}
\oauthor{\bsnm{Vogel}, \binits{A.}},
\oauthor{\bsnm{Henning}, \binits{S.}},
\oauthor{\bsnm{Perez-Wohlfeil}, \binits{E.}},
\oauthor{\bsnm{Ertl}, \binits{O.}},
\oauthor{\bsnm{Rabiser}, \binits{R.}}:
A comprehensive benchmarking analysis of fault recovery in stream processing frameworks.
arXiv preprint arXiv:2404.06203
(2024)
\end{botherref}
\endbibitem

\bibitem[\protect\citeauthoryear{Vogel et~al.}{2023}]{SEAA2023}
\begin{bchapter}
\bauthor{\bsnm{Vogel}, \binits{A.}},
\bauthor{\bsnm{Henning}, \binits{S.}},
\bauthor{\bsnm{Ertl}, \binits{O.}},
\bauthor{\bsnm{Rabiser}, \binits{R.}}:
\bctitle{A systematic mapping of performance in distributed stream processing systems}.
In: \bbtitle{Euromicro Conference on Software Engineering and Advanced Applications},
pp. \bfpage{293}--\blpage{300}
(\byear{2023}).
\bcomment{IEEE}
\end{bchapter}
\endbibitem

\bibitem[\protect\citeauthoryear{Henning et~al.}{2024}]{ICPE2024}
\begin{bchapter}
\bauthor{\bsnm{Henning}, \binits{S.}},
\bauthor{\bsnm{Vogel}, \binits{A.}},
\bauthor{\bsnm{Leichtfried}, \binits{M.}},
\bauthor{\bsnm{Ertl}, \binits{O.}},
\bauthor{\bsnm{Rabiser}, \binits{R.}}:
\bctitle{Shufflebench: A benchmark for large-scale data shuffling operations with distributed stream processing frameworks}.
In: \bbtitle{Proceedings of the 15th ACM/SPEC International Conference on Performance Engineering},
p. \bfpage{12}
(\byear{2024}).
\bcomment{In press}
\end{bchapter}
\endbibitem

\bibitem[\protect\citeauthoryear{Herodotou et~al.}{2020}]{Herodotou2020}
\begin{barticle}
\bauthor{\bsnm{Herodotou}, \binits{H.}},
\bauthor{\bsnm{Chen}, \binits{Y.}},
\bauthor{\bsnm{Lu}, \binits{J.}}:
\batitle{A survey on automatic parameter tuning for big data processing systems}.
\bjtitle{ACM Comput Surv}
\bvolume{53}(\bissue{2}),
\bfpage{1}--\blpage{37}
(\byear{2020})
\end{barticle}
\endbibitem

\bibitem[\protect\citeauthoryear{Vogel et~al.}{2022}]{Vogel2022}
\begin{botherref}
\oauthor{\bsnm{Vogel}, \binits{A.}},
\oauthor{\bsnm{Griebler}, \binits{D.}},
\oauthor{\bsnm{Danelutto}, \binits{M.}},
\oauthor{\bsnm{Fernandes}, \binits{L.G.}}:
Self-adaptation on parallel stream processing: {A} systematic review.
{Concurr Comp}
\textbf{34}(6)
(2022)
\end{botherref}
\endbibitem

\bibitem[\protect\citeauthoryear{Fragkoulis et~al.}{2024}]{Fragkoulis2023}
\begin{barticle}
\bauthor{\bsnm{Fragkoulis}, \binits{M.}},
\bauthor{\bsnm{Carbone}, \binits{P.}},
\bauthor{\bsnm{Kalavri}, \binits{V.}},
\bauthor{\bsnm{Katsifodimos}, \binits{A.}}:
\batitle{A survey on the evolution of stream processing systems}.
\bjtitle{VLDB}
\bvolume{33}(\bissue{2}),
\bfpage{507}--\blpage{541}
(\byear{2024})
\end{barticle}
\endbibitem

\bibitem[\protect\citeauthoryear{Sax et~al.}{2018}]{Sax2018}
\begin{bchapter}
\bauthor{\bsnm{Sax}, \binits{M.J.}},
\bauthor{\bsnm{Wang}, \binits{G.}},
\bauthor{\bsnm{Weidlich}, \binits{M.}},
\bauthor{\bsnm{Freytag}, \binits{J.-C.}}:
\bctitle{Streams and tables: Two sides of the same coin}.
In: \bbtitle{International Workshop on Real-Time Business Intelligence and Analytics}.
\bpublisher{Association for Computing Machinery},
\blocation{New York, USA}
(\byear{2018})
\end{bchapter}
\endbibitem

\bibitem[\protect\citeauthoryear{Dean and Ghemawat}{2008}]{Dean2008}
\begin{barticle}
\bauthor{\bsnm{Dean}, \binits{J.}},
\bauthor{\bsnm{Ghemawat}, \binits{S.}}:
\batitle{{MapReduce}: Simplified data processing on large clusters}.
\bjtitle{Communications of the ACM}
\bvolume{51}(\bissue{1}),
\bfpage{107}--\blpage{113}
(\byear{2008})
\end{barticle}
\endbibitem

\bibitem[\protect\citeauthoryear{van Dongen and van~den Poel}{2021}]{vanDongen2021a}
\begin{barticle}
\bauthor{\bsnm{Dongen}, \binits{G.}},
\bauthor{\bsnm{Poel}, \binits{D.}}:
\batitle{A performance analysis of fault recovery in stream processing frameworks}.
\bjtitle{IEEE Access}
\bvolume{9},
\bfpage{93745}--\blpage{93763}
(\byear{2021})
\end{barticle}
\endbibitem

\bibitem[\protect\citeauthoryear{Wang et~al.}{2022}]{Wang2022}
\begin{barticle}
\bauthor{\bsnm{Wang}, \binits{X.}},
\bauthor{\bsnm{Zhang}, \binits{C.}},
\bauthor{\bsnm{Fang}, \binits{J.}},
\bauthor{\bsnm{Zhang}, \binits{R.}},
\bauthor{\bsnm{Qian}, \binits{W.}},
\bauthor{\bsnm{Zhou}, \binits{A.}}:
\batitle{A comprehensive study on fault tolerance in stream processing systems}.
\bjtitle{{Front Comput Sci}}
\bvolume{16}(\bissue{2}),
\bfpage{162603}
(\byear{2022})
\end{barticle}
\endbibitem

\bibitem[\protect\citeauthoryear{Henning and Hasselbring}{2022}]{EMSE2022}
\begin{botherref}
\oauthor{\bsnm{Henning}, \binits{S.}},
\oauthor{\bsnm{Hasselbring}, \binits{W.}}:
A configurable method for benchmarking scalability of cloud-native applications.
Empirical Software Engineering
\textbf{27}(6)
(2022)
\end{botherref}
\endbibitem

\bibitem[\protect\citeauthoryear{Basiri et~al.}{2016}]{Basiri2016}
\begin{barticle}
\bauthor{\bsnm{Basiri}, \binits{A.}},
\bauthor{\bsnm{Behnam}, \binits{N.}},
\bauthor{\bsnm{Rooij}, \binits{R.}},
\bauthor{\bsnm{Hochstein}, \binits{L.}},
\bauthor{\bsnm{Kosewski}, \binits{L.}},
\bauthor{\bsnm{Reynolds}, \binits{J.}},
\bauthor{\bsnm{Rosenthal}, \binits{C.}}:
\batitle{Chaos engineering}.
\bjtitle{IEEE Software}
\bvolume{33}(\bissue{3}),
\bfpage{35}--\blpage{41}
(\byear{2016})
\end{barticle}
\endbibitem

\bibitem[\protect\citeauthoryear{Guo et~al.}{2021}]{guo2021}
\begin{barticle}
\bauthor{\bsnm{Guo}, \binits{Y.}},
\bauthor{\bsnm{Shan}, \binits{H.}},
\bauthor{\bsnm{Huang}, \binits{S.}},
\bauthor{\bsnm{Hwang}, \binits{K.}},
\bauthor{\bsnm{Fan}, \binits{J.}},
\bauthor{\bsnm{Yu}, \binits{Z.}}:
\batitle{{GML}: efficiently auto-tuning flink's configurations via guided machine learning}.
\bjtitle{IEEE Transactions on Parallel and Distributed Systems}
\bvolume{32}(\bissue{12}),
\bfpage{2921}--\blpage{2935}
(\byear{2021})
\end{barticle}
\endbibitem

\bibitem[\protect\citeauthoryear{Li et~al.}{2023}]{li2023}
\begin{barticle}
\bauthor{\bsnm{Li}, \binits{Y.}},
\bauthor{\bsnm{Jiang}, \binits{H.}},
\bauthor{\bsnm{Shen}, \binits{Y.}},
\bauthor{\bsnm{Fang}, \binits{Y.}},
\bauthor{\bsnm{Yang}, \binits{X.}},
\bauthor{\bsnm{Huang}, \binits{D.}},
\bauthor{\bsnm{Zhang}, \binits{X.}},
\bauthor{\bsnm{Zhang}, \binits{W.}},
\bauthor{\bsnm{Zhang}, \binits{C.}},
\bauthor{\bsnm{Chen}, \binits{P.}}, \betal:
\batitle{Towards general and efficient online tuning for spark}.
\bjtitle{Proceedings of the VLDB Endowment}
\bvolume{16}(\bissue{12}),
\bfpage{3570}--\blpage{3583}
(\byear{2023})
\end{barticle}
\endbibitem

\bibitem[\protect\citeauthoryear{Zhang et~al.}{2024}]{zhang2024}
\begin{botherref}
\oauthor{\bsnm{Zhang}, \binits{B.}},
\oauthor{\bsnm{Liu}, \binits{Z.}},
\oauthor{\bsnm{Cherry}, \binits{C.}},
\oauthor{\bsnm{Firat}, \binits{O.}}:
{When Scaling Meets LLM Finetuning: The Effect of Data, Model and Finetuning Method}.
arXiv preprint arXiv:2402.17193
(2024)
\end{botherref}
\endbibitem

\bibitem[\protect\citeauthoryear{Geldenhuys et~al.}{2022}]{geldenhuys2022}
\begin{bchapter}
\bauthor{\bsnm{Geldenhuys}, \binits{M.K.}},
\bauthor{\bsnm{Pfister}, \binits{B.J.}},
\bauthor{\bsnm{Scheinert}, \binits{D.}},
\bauthor{\bsnm{Thamsen}, \binits{L.}},
\bauthor{\bsnm{Kao}, \binits{O.}}:
\bctitle{Khaos: Dynamically optimizing checkpointing for dependable distributed stream processing}.
In: \bbtitle{FedCSIS},
pp. \bfpage{553}--\blpage{561}
(\byear{2022}).
\bcomment{IEEE}
\end{bchapter}
\endbibitem

\bibitem[\protect\citeauthoryear{Andrade et~al.}{2021}]{andrade2021}
\begin{bchapter}
\bauthor{\bsnm{Andrade}, \binits{G.}},
\bauthor{\bsnm{Griebler}, \binits{D.}},
\bauthor{\bsnm{Santos}, \binits{R.}},
\bauthor{\bsnm{Danelutto}, \binits{M.}},
\bauthor{\bsnm{Fernandes}, \binits{L.G.}}:
\bctitle{Assessing coding metrics for parallel programming of stream processing programs on multi-cores}.
In: \bbtitle{Euromicro Conference on Software Engineering and Advanced Applications},
pp. \bfpage{291}--\blpage{295}
(\byear{2021}).
\bcomment{IEEE}
\end{bchapter}
\endbibitem

\end{thebibliography}

\end{document}